\documentclass[12pt,preprint]{aastex}

\newcommand{\eqref}[1]{(\ref{#1})}

\newcommand{\In}{{I_{n}}}
\newcommand{\Jn}{{J_{n}}}

\newcommand{\Ink}{{I_{k,n}}}
\newcommand{\Jnk}{{J_{k,n}}}

\newcommand{\Jtnj}{{\tilde{J}_{j,n}}}
\newcommand{\Jtnjp}{{\tilde{J}_{j+1,n}}}
\newcommand{\Jtnjm}{{\tilde{J}_{j-1,n}}}

\newcommand{\Jtnn}{{\tilde{J}_{n/2,n}}}
\newcommand{\Jtnnpm}{{\tilde{J}_{n/2 \pm 1,n}}}

\newcommand{\ds}{\displaystyle}

\newcommand{\sumj}{{\sum_{k \ne j}}}
\newcommand{\sumn}{{\sum_{k \ne n}}}
\newcommand{\sumjn}{{\sum_{k \ne j,n}}}
\newcommand{\sumzero}{{\sum_{k=1}^{n-1}}}
\newcommand{\zkj}{{\xi_{k,j}}}
\newcommand{\dzkj}{{\delta \xi_{k,j}}}
\newcommand{\ztwokj}{{\xi_{k,j}^2}}
\newcommand{\zbarkj}{{\bar{\xi}_{k,j}}}
\newcommand{\dzbarkj}{{\delta \zbarkj}}
\newcommand{\tkj}{{\theta_{k-j}}}

\newcommand{\tkn}{{\theta_{k}}}

\shorttitle{Stability of Ring Systems}
\shortauthors{R.J. Vanderbei and E. Kolemen}

\begin{document}

\title{Linear Stability of Ring Systems}

\author{Robert J. Vanderbei}
\affil{Operations Research and Financial Engineering, Princeton University}
\email{rvdb@princeton.edu}

\author{Egemen Kolemen}
\affil{Mechanical and Aerospace Engineering, Princeton University}
\email{ekolemen@princeton.edu}

\begin{abstract}
%
We give a self-contained modern linear stability analysis of a system of $n$
equal mass bodies in circular orbit about a single more massive body.
Starting with the mathematical description of
the dynamics of the system, we form the linear approximation, compute all
of the eigenvalues of the linear stability matrix, and
finally derive inequalities that guarantee that none of these eigenvalues
have positive real part.  In the end, we rederive the result that
J.C. Maxwell found for large $n$ in his seminal paper on the nature and
stability of Saturn's rings, which was published 150 years ago.
In addition, we identify the exact matrix that defines the
linearized system even when $n$ is not large.  This matrix is then
investigated numerically (by computer) to find stability inequalities.
Furthermore, using properties of circulant
matrices, the eigenvalues of the large $4n \times 4n$ matrix can be computed
by solving $n$ quartic equations, which further facilitates the investigation
of stability.  Finally, we have implemented an $n$-body
simulator and we verify that the
threshold mass ratios that we derived mathematically or numerically do indeed
identify the threshold between stability and instability.
Throughout the paper we consider only the planar $n$-body problem so that the
analysis can be carried out
purely in complex notation, which makes the equations and derivations more
compact, more elegant and therefore, we hope, more transparent.
The result is a fresh analysis that shows that these systems are always
unstable for $2 \le n \le 6$ and for $n>6$ they are stable provided that the central
mass is massive enough.  We give an explicit formula for this mass-ratio
threshold.
\end{abstract}

\keywords{planets: rings}

\section{Introduction}

One hundred and fifty years ago, \citet{Max59} was awarded the
prestigious Adam's prize for a seminal
paper on the stability of Saturn's rings.  At that time, neither the structure
nor the composition of the rings was known.  Hence, Maxwell considered various
scenarios such as the possibility that the rings were solid or liquid annuli
or a myriad of small boulders.  As a key part of this last
possibility, Maxwell studied the case of $n$ equal-mass bodies orbiting
Saturn at a common radius and uniformly distributed about a circle of this
radius.  He concluded that, for large $n$, the ring ought to be stable
provided that the following inequality is satisfied:
\[
    \mbox{mass(Rings)} \le 2.298 \mbox{mass(Saturn)}/n^2 .
\]
The mathematical analysis that leads to this result has been scrutenized,
validated, and generalized by a number of mathematicians over the years.

We summarize briefly some of the key historical developments.
\citet{1889Tisserand} derived the same stability criterion using an analysis
where he assumed that the ring has no effect on Saturn and that the highest
vibration mode of the system controls stability.  More recently,
\citet{1986A&A...161..403W} used the theory of density waves to show that
Maxwell's results are correct in the limit as $n$ goes to infinity.
\citet{1935RSPTA.234..145P} reformulated the stability problem so that it
takes into account the effect of the rings on the central body.  He proved
that, for $n \le 6$, the system is unconditionally unstable.  Inspired by this
work, \citet{1988A&A...205..309S} studied coorbital formations of $n$
satellites for small values of $n$ where the satellites are not distributed
uniformly around the central body.  They showed that there are some stable
asymmetric formations (such as the well-known case of a pair of ring bodies in
L4/L5 position relative to each other---i.e., one leading the other by
$60\deg$).  Finally, \citet{1991CeMDA..51...83S} extended the analysis of
Pendse to find the stability criterion as a function of the number of
satellites when $n$ is small.   The resulting threshold depends on $n$ but for
$n \ge 7$, it deviates only a small amount from the asymptotically derived
value.


In this paper, we give a self-contained
modern linear stability analysis of a system of equal
mass bodies in circular orbit about a single more massive body.  We start with
the mathematical description of
the dynamics of the system. We then form the linear approximation, compute all
of the eigenvalues of the matrix defining the linear approximation, and
finally we derive inequalities that guarantee that none of these eigenvalues
have positive real part.  In the end, we get exactly the same result that
Maxwell found for large $n$.
But, in addition, we identify the exact matrix that defines the
linearized system even when $n$ is not large.  This matrix can then be
investigated numerically to find stability inequalities even in
cases where $n$ is not large.  Furthermore, using properties of circulant
matrices, the eigenvalues of the large $4n \times 4n$ matrix can be computed
by solving $n$ quartic equations, which further facilitates the investigation.
Finally, we have implemented an $n$-body
simulator based on a leap-frog integrator (see \cite{ST94,HMM94})
and we verify that the
threshold mass ratios that we derived mathematically or numerically do indeed
identify the threshold between stability and instability.

Throughout the paper we consider only the planar $n$-body problem.  That is,
we ignore any instabilities that might arise due to out-of-plane
perturbations.  Maxwell claimed, and others have confirmed, that these
out-of-plane perturbations are less destabilizing than in-plane ones and hence
our analysis, while not fully general, does get to the right answer.  Our main
reason for wishing to restrict to the planar case is that we can then work
in the complex plane and our entire analysis can be carried out
purely in complex notation, which makes the equations and derivations more
compact, more elegant and therefore, we hope, more transparent.

\section{Equally-Spaced, Equal-Mass Bodies in a Circular 
Ring About a Massive Body}
Consider the multibody problem consisting of one large central body, say
Saturn, having mass $M$ and $n$ small bodies, such as boulders, each of
mass $m$ orbiting the large body in circular orbits uniformly spaced in a ring
of radius $r$.  Indices $0$ to $n-1$ will be used to denote the ring masses
and index $n$ will be used for Saturn.  Throughout the paper we assume that
$n \ge 2$.  For the case $n=1$, Lagrange proved that the system
is stable for all mass ratios $m/M$.

The purpose of this section is to show that such a ring exists as a solution
to Newton's law of gravitation.  In particular, we derive the relationship
between the angular velocity $\omega$ of the ring particles and their radius
$r$ from the central mass.
We assume all bodies lie in a plane and therefore complex-variable notation is
convenient.
So, with $i = \sqrt{-1}$ and $z = x + iy$, we can write the equilibrium
solution for $j=0,1,\ldots,n-1$, as
\begin{equation} \label{1}
    z_j = r e^{i(\omega t + 2 \pi j/n)}
\end{equation}
and
\begin{equation} \label{2}
    z_n = 0.
\end{equation}
By symmetry (and exploiting our assumption that $n \ge 2$), force is balanced on Saturn itself.
Now consider the ring bodies.
Differentiating \eqref{1}, we see that
\begin{equation} \label{6}
    \ddot{z}_j = - \omega^2 z_j .
\end{equation}
From Newton's law of gravity we have that
\begin{equation} \label{3}
    \ddot{z}_j = - GM \frac{z_j-z_n}{|z_j-z_n|^3}
               + \sumjn Gm \frac{z_k - z_j}{|z_k - z_j|^3}.
\end{equation}
Equations \eqref{6} and \eqref{3} allow us to determine $\omega$, which is our
first order of business.
By symmetry it suffices to consider $j=0$.
It is easy to check that
\begin{equation} \label{4}
    z_k - z_0 = re^{i \omega t} e^{\pi i k/n} 2 i \sin(\pi k/n)
\end{equation}
and hence that
\begin{equation} \label{5}
    |z_k - z_0| = 2 r \sin(\pi k/n) .
\end{equation}
Substituting \eqref{4} and \eqref{5} into \eqref{3} and equating this with
\eqref{6}, we see that
\begin{eqnarray} \label{7}
    -\omega^2 & = & -\frac{GM}{r^3}
                + \sum_{k=1}^{n-1} \frac{Gm}{4r^3}
            \frac{i e^{\pi i k/n}}{\sin^2(\pi k/n)} \\
              & = & -\frac{GM}{r^3}
                    - \frac{Gm}{4r^3}
              \sum_{k=1}^{n-1}
            \frac{1}{\sin(\pi k/n)}
                    +i\frac{Gm}{4r^3}
              \sum_{k=1}^{n-1}
            \frac{\cos(\pi k/n)}{\sin^2(\pi k/n)} .
            \label{8}
\end{eqnarray}
It is easy to check that the summation in the imaginary part on the right
vanishes.  Hence,
\begin{equation} \label{9}
    \omega^2 = \frac{GM}{r^3} + \frac{Gm}{r^3} \In
\end{equation}
where
\begin{equation} \label{10}
    \In = \frac{1}{4} \sum_{k=1}^{n-1} \frac{1}{\sin(\pi k/n)} .
\end{equation}
With this choice of $\omega$, the trajectories given by \eqref{1} and \eqref{2}
satisfy Newton's law of gravitation.

\section{First-Order Stability}

In order to carry out a stability analysis, we need to counter-rotate the
system so that all bodies remain at rest.  We then perturb the system slightly
and analyze the result.

A counter-rotated system would be given by
\[
    e^{- i \omega t} z_j(t) = r e^{2 \pi i j/n} = z_j(0).
\]
In such a rotating frame of reference, each body remains fixed at its initial
point.  It turns out to be better to rotate the different bodies different
amounts so that every ring body is repositioned to lie on the $x$-axis.  In
other words, for $j=0,\ldots,n-1,n$, we define
\begin{equation} \label{11}
    w_j
    = u_j + i v_j
    = e^{- i (\omega t + 2 \pi j/n)} z_j
    .
\end{equation}
The advantage of repositioning every ring body to the positive real axis
is that
perturbations in the real part for any ring body represent radial perturbations
whereas perturbations in the imaginary part represent azimuthal perturbations.
A simple counter-rotation does not provide such a clear distinction
between the two types of perturbations (and the associated stability matrix
fails to have the circulant property that is crucial to all later analysis).

Differentiating \eqref{11} twice, we get
\begin{equation} \label{12}
    \ddot{w}_j = \omega^2 w_j
               - 2 i \omega \dot{w}_j
               + e^{- i (\omega t + 2 \pi j/n)} \ddot{z}_j
           .
\end{equation}
From Newton's law of gravity, we see that
\begin{equation} \label{13}
    \ddot{w}_j = \omega^2 w_j
               - 2 i \omega \dot{w}_j
               + \sumj Gm_k \frac{\zkj}{|\zkj|^3}
           ,
\end{equation}
where
\begin{equation} \label{202}
    m_k = \cases{
                 m, & for $ k = 0,1,\ldots,n-1$, \cr
                 M, & for $ k = n$, \cr
          }
\end{equation}
\begin{equation} \label{200}
    \zkj = e^{i \tkj} w_k - w_j
\end{equation}
and
\begin{equation} \label{201}
    \theta_k = 2 \pi k / n .
\end{equation}

Let $\delta w_j(t)$ denote variations
about the fixed point given by
\begin{equation} \label{211}
    w_j \equiv \cases{
                   r, & for $j = 0,1,\ldots,n-1$, \cr
                   0, & for $j = n$. \cr
               }
\end{equation}
We compute a linear
approximation to the differential equation describing the evolution of such a
perturbation.  Applying the quotient, chain, and product rules as
needed, we get
\begin{eqnarray}
    \ddot{\delta w}_j & = &
                 \omega^2 {\delta w}_j
           - 2 i \omega \dot{\delta w}_j
                   + \sumj Gm_k
            \frac{|\zkj|^3 \dzkj - \zkj \frac{3}{2}|\zkj|
                 (\zkj \dzbarkj + \zbarkj \dzkj)}{
              |\zkj|^6
                 }
             \nonumber \\
                      & = &
                 \omega^2 {\delta w}_j
           - 2 i \omega \dot{\delta w}_j
                   - \frac{1}{2} \sumj Gm_k
            \frac{|\zkj|^2 \dzkj + 3 \ztwokj \dzbarkj}{
              |\zkj|^5
                 } , \label{210}
\end{eqnarray}
where 
\begin{eqnarray*}
    \dzkj    &=& e^{i \tkj} \delta w_k - \delta w_j \\[0.2in]
    \dzbarkj &=& e^{-i \tkj} \delta \bar{w}_k - \delta \bar{w}_j .
\end{eqnarray*}

The next step is to use \eqref{200} to re-express the $\zkj$'s in terms of the
$w_k$'s and the $w_j$'s and then to substitute in
the particular solution given by \eqref{211}.
%
%
Consider the case where $j < n$.  In this case we have
\[
    \zkj = \cases{
               r (e^{i \tkj} - 1), & for $k < n$, \cr
              -r,                  & for $k = n$ \cr
           }
\]
and therefore
\[
    |\zkj| = \cases{
                2r \sin (|\tkj|/2), & for $ k < n$, \cr
                 r,                 & for $ k = n$. \cr
             }
\]
Substituting these into \eqref{210} and simplifying, we get
\begin{eqnarray}
    \ddot{\delta w}_j &=&
                 \omega^2 {\delta w}_j
           - 2 i \omega \dot{\delta w}_j
           - \frac{GM}{2r^3}
              (e^{-i\theta_j} \delta w_n
              + 3 e^{i\theta_j} \delta \bar{w}_n)
           + \frac{GM}{2r^3}
              (\delta w_j + 3 \delta \bar{w}_j)
           \nonumber \\[0.2in]
        && \quad - \frac{Gm}{2r^3} \; \frac{1}{8}
           \sumjn
           \frac{e^{i\tkj}\delta w_k - \delta w_j
             -3e^{i\tkj}(e^{-i\tkj}\delta\bar{w}_k - \delta\bar{w}_j)}{
              \sin^3 (|\tkj|/2)
              }
           . \label{215}
\end{eqnarray}

\section{Choice of Coordinate System}
Without loss of generality, we can choose our coordinate system so that the
center of mass remains fixed at the origin.  Having done that, 
the perturbations $\delta w_n$ and $\delta \bar{w}_n$ can be computed
explicitly in terms of the other perturbations.  Indeed, conservation of
momentum implies that
\[
    m \sumn \delta z_k + M \delta z_n = 0.
\]
Hence,
\[
    \delta z_n = - \frac{m}{M} \sumn \delta z_k .
\]
From the definition \eqref{11} of the $w_k$'s in terms of the $z_k$'s, it then
follows that
\[
    e^{-i \theta_j} \delta w_n = - \frac{m}{M} \sumn e^{i \tkj} \delta w_k .
\]
Making this substitution for $e^{-i \theta_j} \delta w_n$
and an analogous substitution for $e^{i \theta_j} \delta \bar{w}_n$
in \eqref{215}, we see that
\begin{eqnarray}
    \ddot{\delta w}_j &=&
                 \omega^2 {\delta w}_j
           - 2 i \omega \dot{\delta w}_j
           + \frac{Gm}{2r^3} \sumn
              \left(e^{i\theta_{k-j}} \delta w_k
              + 3 e^{-i\theta_{k-j}} \delta \bar{w}_k \right)
           + \frac{GM}{2r^3}
              (\delta w_j + 3 \delta \bar{w}_j)
           \nonumber \\[0.2in]
        && \quad - \frac{Gm}{2r^3} \; \frac{1}{8}
           \sumjn
           \frac{e^{i\tkj}\delta w_k - \delta w_j
             -3e^{i\tkj}(e^{-i\tkj}\delta\bar{w}_k - \delta\bar{w}_j)}{
              \sin^3 (|\tkj|/2)
              } .
            \label{216}
\end{eqnarray}

\section{Circulant Matrix}

Switching to matrix notation,
let $W_j$ denote a shorthand for the column vector
$\left[\begin{array}{cc} w_j & \bar{w}_j \end{array}\right]'$.
In this notation, we see that \eqref{216} together with its conjugates
can be written as
\begin{equation} \label{42}
    \frac{d}{dt}
    \left[ \begin{array}{c}
        {\delta W}_0 \\
        {\delta W}_1 \\
        \vdots       \\
        {\delta W}_{n-1} \\
        \dot{\delta W}_0 \\
        \dot{\delta W}_1 \\
        \vdots           \\
        \dot{\delta W}_{n-1}
    \end{array} \right]
    \approx
    \left[ \begin{array}{cccc|cccc}
      &   &   &   &   I &   &        &    \\
      &   &   &   &     & I &        &    \\
      &   &   &   &     &   & \ddots &    \\
      &   &   &   &     &   &        & I  \\
    \hline
    D       & N_1  &\cdots&N_{n-1}&  \Omega &         &      &   \\
    N_{n-1} & D    &\cdots&N_{n-2}&         & \Omega  &      &   \\
    \vdots  &\vdots&      &\vdots &         &         &\ddots&   \\
    N_1     & N_2  &\cdots& D     &         &         &      & \Omega
    \end{array} \right]
    \left[ \begin{array}{c}
        {\delta W}_0 \\
        {\delta W}_1 \\
        \vdots       \\
        {\delta W}_{n-1} \\
        \dot{\delta W}_0 \\
        \dot{\delta W}_1 \\
        \vdots           \\
        \dot{\delta W}_{n-1}
    \end{array} \right] ,
\end{equation}
where $D$, $\Omega$, and the $N_k$'s are $2 \times 2$ complex matrices given by
\begin{eqnarray*}
    D &=&
    \frac{3}{2} \omega^2
    \left[ \begin{array}{cc}
        1 & 1 \\ 1 & 1
    \end{array} \right]
    +
    \frac{Gm}{2 r^3}
    \left[ \begin{array}{cc}
        \ds 1 - \In + \Jn/2 & \ds 3 - 3\Jn/2 \\
    \ds 3 - 3\Jn/2     & \ds 1 - \In + \Jn/2
    \end{array} \right]
    \\[0.2in]
    N_k &=&
    \frac{Gm}{2r^3}
    \left[ \begin{array}{cc}
        \ds e^{i \theta_k} \left( 1 - \Jnk/2 \right)
    &
        \ds 3 e^{-i \theta_k} + 3\Jnk/2
    \\
        \ds 3 e^{i \theta_k} + 3\Jnk/2
    &
        \ds e^{-i \theta_k} \left( 1 - \Jnk/2 \right)
    \end{array} \right]
    \\[0.2in]
    \Omega &=&
    2 i \omega
    \left[ \begin{array}{cc}
        -1 & 0 \\ 0 & 1
    \end{array} \right]
    ,
\end{eqnarray*}
and where
\begin{eqnarray*}
    \Ink &=& \frac{1}{4 \sin\left(\pi |k|/n\right)} \\
    \Jnk &=& \frac{1}{4 \sin^3\left(\pi |k|/n\right)} 
\end{eqnarray*}
and
\begin{eqnarray}
    \In &=& \sumzero \Ink
            \; \approx \; \frac{1}{2 \pi} \; n \sum_{k=1}^{(n-1)/2} \frac{1}{k}
            \; \approx \; \frac{1}{2 \pi} \; n \log (n/2) \label{310} \\[0.2in]
    \Jn &=& \sumzero \Jnk \; \approx \; \frac{1}{2 \pi^3} n^3
                                         \sum_{k=1}^{\infty} \frac{1}{k^3}
                     \; =  \frac{n^3}{2 \pi^3} \;  \zeta(3) = 0.01938 \; n^3.
                     \label{311}
\end{eqnarray}
Here, the symbol $\approx$ is used to indicate asymptotic agreement.  That
is, $a_n \approx b_n$ means that $a_n/b_n \rightarrow 1$ as $n \rightarrow
\infty$ and $\zeta(3)$ denotes the value of the Riemann zeta function at $3$.  
This constant is known as {\em Ap\'{e}ry's constant}
(see, e.g., \cite{Arf85}).  
%
%

Finally, note that in deriving \eqref{42} from \eqref{216} we
have made repeated use of the following identity
\[
    \sum_{k=1}^{n-1} \frac{e^{i \theta_k}}{\sin^3 |\theta_k|/2}
    = 4 \Jn - 8 \In .
\]

Let $A$ denote the matrix in \eqref{42}.
We need to find the eigenvalues of $A$ and derive necessary and sufficient
conditions under which none of them have a positive real part.
At this point we could resort to numerical computation to bracket a threshold
for stability by doing a binary search to find the largest value of $m/M$ for
which none of the eigenvalues have positive real part.  We did such a search
for some values of $n$.  The results are shown in Table \ref{tab1}.

The eigenvalues are complex numbers for which there are nontrivial solutions
to
\begin{equation} \label{101}
    \left[ \begin{array}{cccc|cccc}
      &   &   &   & I &   &   &   \\
      &   &   &   &   & I &   &   \\
      &   &   &   &   &   & \ddots &   \\
      &   &   &   &   &   &   & I \\
    \hline
    D       & N_1  &\cdots&N_{n-1}& \Omega &         &      &   \\
    N_{n-1} & D    &\cdots&N_{n-2}&        & \Omega  &      &   \\
    \vdots  &\vdots&      &\vdots &        &         &\ddots&   \\
    N_1     & N_2  &\cdots& D     &        &         &      & \Omega
    \end{array} \right]
    \left[ \begin{array}{c}
        {\delta W}_0 \\
        {\delta W}_1 \\
        \vdots       \\
        {\delta W}_{n-1} \\
        \dot{\delta W}_0 \\
        \dot{\delta W}_1 \\
        \vdots           \\
        \dot{\delta W}_{n-1}
    \end{array} \right]
    =
    \lambda
    \left[ \begin{array}{c}
        {\delta W}_0 \\
        {\delta W}_1 \\
        \vdots       \\
        {\delta W}_{n-1} \\
        \dot{\delta W}_0 \\
        \dot{\delta W}_1 \\
        \vdots           \\
        \dot{\delta W}_{n-1}
    \end{array} \right] .
\end{equation}
The first group of equations (above the line) can be used to eliminate the
``derivative'' variables from the second set.  That is,
\[
    \left[ \begin{array}{c}
        \dot{\delta W}_0 \\
        \dot{\delta W}_1 \\
        \vdots           \\
        \dot{\delta W}_{n-1}
    \end{array} \right]
    =
    \lambda
    \left[ \begin{array}{c}
        {\delta W}_0 \\
        {\delta W}_1 \\
        \vdots       \\
        {\delta W}_{n-1} \\
    \end{array} \right]
\]
and therefore
\begin{equation} \label{100}
    \left[ \begin{array}{cccc}
    D       & N_1  &\cdots&N_{n-1} \\
    N_{n-1} & D    &\cdots&N_{n-2} \\
    \vdots  &\vdots&      &\vdots  \\
    N_1     & N_2  &\cdots& D      \\
    \end{array} \right]
    \left[ \begin{array}{c}
        {\delta W}_0 \\
        {\delta W}_1 \\
        \vdots       \\
        {\delta W}_{n-1} \\
    \end{array} \right]
    +
    \lambda
    \left[ \begin{array}{cccc}
    \Omega &         &      &   \\
           & \Omega  &      &   \\
           &         &\ddots&   \\
           &         &      & \Omega
    \end{array} \right]
    \left[ \begin{array}{c}
        {\delta W}_0 \\
        {\delta W}_1 \\
        \vdots       \\
        {\delta W}_{n-1} \\
    \end{array} \right]
    =
    \lambda^2
    \left[ \begin{array}{c}
        {\delta W}_0 \\
        {\delta W}_1 \\
        \vdots       \\
        {\delta W}_{n-1} \\
    \end{array} \right] .
\end{equation}
The matrix on the left-hand side is called a {\em block circulant matrix}.  
Much is known about such matrices.  In particular, it is easy to find the
eigenvectors of such matrices.  For general properties of block
circulant matrices, see \cite{Tee05}.

Let $\rho$ denote an $n$-th root
of unity (i.e., $\rho = e^{2 \pi i j/n}$ for some
$j=0,1,\ldots,n-1$) and let $\xi$ be an arbitrary complex
$2$-vector. We look for solutions of the form
\[
    \left[ \begin{array}{c}
        {\delta W}_0 \\
        {\delta W}_1 \\
        \vdots       \\
        {\delta W}_{n-1} \\
    \end{array} \right]
    =
    \left[ \begin{array}{c}
        \xi \\
        \rho \xi \\
        \vdots       \\
        \rho^{n-1} \xi
    \end{array} \right] .
\]
Substituting such a guess into \eqref{100}, we see that each of the $n$ rows
reduce to one and the same thing
\[
    \left( D + \rho N_1 + \dots + \rho^{n-1} N_{n-1} \right) \xi
    + \lambda \Omega \xi = \lambda^2 \xi .
\]
There are nontrivial solutions to this $2 \times 2$ system if and only if
\[
    \det( D + \rho N_1 + \dots + \rho^{n-1} N_{n-1}
         + \lambda \Omega - \lambda^2 I)
     = 0.
\]
For each root of unity, $\rho$, there are four values of $\lambda$ that
solve this equation (counting multiplicites).
That makes a total of $4 n$ eigenvalues and therefore provides all eigenvalues
for the full system \eqref{101}.

\section{Explicit Expression for $\sumzero \rho^k N_k$}

In order to compute the eigenvalues, it is essential that we compute $\sumzero
\rho^k N_k$ as explicitly as possible.  To this end, we note the following
reduction and new definition:
\begin{eqnarray}
    \sumzero \rho^k \Jnk
    & = & \frac{1}{4} \sumzero \frac{e^{2\pi ijk/n}}{\sin^3 (\tkn/2)}
    \nonumber \\
    & = & \frac{1}{4} \sumzero \frac{\cos(j\tkn)}{\sin^3 (\tkn/2)} \nonumber \\
    & =: & \Jtnj . \label{300}
\end{eqnarray}
Similarly,
\begin{equation}
    \sumzero \rho^k e^{i \tkn} \Jnk = \Jtnjp \label{301}
\end{equation}
and
\begin{equation}
    \sumzero \rho^k e^{-i \tkn} \Jnk = \Jtnjm . \label{302}
\end{equation}
Also we compute
\begin{equation}
    \sumzero \rho^k e^{i \tkn}
    = \sumzero e^{ij\tkn} {e^{i\tkn}}
    = \sumzero {e^{i(j+1)\tkn}}
    = \cases{
            n-1 & for $j = n-1$ \cr
             -1 & otherwise  \cr
      } \label{303}
\end{equation}
%
and
\begin{equation}
    \sumzero \rho^k e^{-i \tkn}
        = \cases{
            n-1 & for $ j = 1$ \cr
             -1 & otherwise . \cr
          } \label{304}
\end{equation}
Substituting the definition of $N_k$ into
$\sumzero \rho^k N_k$ and making use of \eqref{300}-\eqref{304}, we get
\begin{equation}
    \sumzero \rho^k N_k
    =
    \frac{Gm}{2r^3}
    \left[\begin{array}{cc}
        -1 + n\delta_{j=n-1} - \frac{1}{2} \Jtnjp &
    -3 + 3n\delta_{j=1} + \frac{3}{2} \Jtnj \\
    -3 + 3n\delta_{j=n-1} + \frac{3}{2} \Jtnj &
        -1 + n\delta_{j=1} - \frac{1}{2} \Jtnjm
    \end{array}\right],
    \label{305}
\end{equation}
where $\delta_{j=k}$ denotes the Kronecker delta (i.e., one when $j=k$
and zero otherwise).

\section{Solving
$\det\left( D+\sumzero \rho^k N_k + \lambda \Omega - \lambda^2 I \right) = 0.$}

Assembling the results from the previous sections, we see that
\begin{eqnarray}
    &&
    D + \sum_{k=1}^{n-1} \rho^k N_k
              + \lambda \Omega - \lambda^2 I
    \nonumber \\[0.2in]
    && \qquad
    =
    \left[\begin{array}{cc}
        \frac{3}{2}\omega^2 + \frac{1}{2}\alpha_{j+1}^2 - \beta^2 -
        2i\omega\lambda -\lambda^2
    &
        \frac{3}{2}\omega^2 - \frac{3}{2}\alpha_j^2
    \\
        \frac{3}{2}\omega^2 - \frac{3}{2}\alpha_j^2
    &
        \frac{3}{2}\omega^2 + \frac{1}{2}\alpha_{j-1}^2 - \beta^2 +
        2i\omega\lambda -\lambda^2
    \end{array}\right]
    \nonumber \\[0.2in]
    && \qquad \qquad \qquad \qquad
    +\frac{Gm}{2r^3}
    \left[\begin{array}{cc}
        n\delta_{j=n-1} & 3n\delta_{j=1} \\
    3n\delta_{j=n-1} & n\delta_{j=1}
    \end{array}\right],
    \label{306}
\end{eqnarray}
where $\alpha_j^2$ and $\beta^2$ are shorthands for the expressions
\[
    \alpha_j^2 = \frac{Gm}{2r^3} (\Jn - \Jtnj) \ge 0,
\]
and
\[
    \beta^2 = \frac{Gm}{2r^3} \In \ge 0 .
\]
and, as a reminder, $\In$ and $\Jn$ are defined by \eqref{310} and \eqref{311},
respectively, whereas $\Jtnj$ is defined by \eqref{300}.

It turns out in our subsequent analysis that the root of unity given by
$j = n/2$ is the most critical one for stability, at least for $n \ge 7$.
For $n=2,\ldots,6$ the instability stems from the eigenvectors associated with
$j=1$ and $j=n-1$.
We will analyze the key cases.
But first, we note that the critical $j=n/2$ case corresponds to perturbations
in which every other body is perturbed in the opposite direction.  And, more
importantly, it doesn't matter what the direction of the perturbation is.
That is, if body $0$ is advanced azimuthally, then all of the even-numbered
bodies are advanced azimuthally and all 
of the odd-numbered bodies are retarded by the same
amount.  Similarly, if body $0$ is pushed outward radially, then all of the
even-numbered bodies are also pushed outward whereas the odd-numbered bodies
are pull inward.  Azimuthal and radial perturbations contribute equally to
instability.

\subsection{The Case Where $n$ Is Arbitrary And $j$ Is Neither $1$ Nor $n-1$.}

Assuming that $j$ is neither $1$ nor $n-1$, we see that
\begin{eqnarray}
    \det\left( D+\sumzero \rho^k N_k + \lambda \Omega - \lambda^2 I \right)
    & = &
    \left( \frac{3}{2} \omega^2 + \frac{1}{2}\alpha_{j+1}^2 - \beta^2
        - 2i\omega\lambda - \lambda^2 \right) \nonumber \\
    && \qquad \times
    \left( \frac{3}{2} \omega^2 + \frac{1}{2}\alpha_{j-1}^2 - \beta^2
        + 2i\omega\lambda - \lambda^2 \right) \nonumber \\
    && -
    \left( \frac{3}{2} \omega^2 - \frac{3}{2} \alpha_j^2 \right)^2 .
    \label{400}
\end{eqnarray}
%
Expanding out the products on the right-hand side in \eqref{400},
we get that
\begin{equation}
    \det\left( D+\sumzero \rho^k N_k + \lambda \Omega - \lambda^2 I \right)
    =
    \lambda^4 + A_j \lambda^2 + iB_j \lambda + C_j = 0,
    \label{410}
\end{equation}
where
\begin{eqnarray}
    A_j & = & \omega^2
             - \frac{1}{2} \left( \alpha_{j-1}^2 + \alpha_{j+1}^2 \right)
         + 2 \beta^2 \label{411} \\
    B_j & = & - \omega \left( \alpha_{j-1}^2 - \alpha_{j+1}^2 \right)
            \label{412} \\
    C_j & = & 3 \omega^2 \left( \frac{1}{4} (\alpha_{j-1}^2+\alpha_{j+1}^2)
                  + \frac{3}{2} \alpha_j^2 - \beta^2
                         \right)
           + \left( \frac{1}{4} (\alpha_{j-1}^2 + \alpha_{j+1}^2)
                - \beta^2 \right)^2
    \nonumber \\ &   & \quad
            - \frac{1}{16} (\alpha_{j-1}^2-\alpha_{j+1}^2)^2
                    - \frac{9}{4} \alpha_j^4 . \label{413}
\end{eqnarray}

\subsubsection{The Subcase Where $n$ Is Even And $j = n/2$.}

In the subcase where $n$ is even and $j=n/2$, it is easy to see by symmetry that
$\alpha_{n/2+1} = \alpha_{n/2-1}$.  To emphasize the equality, we will denote
this common value by $\alpha_{n/2 \pm 1}$.  Equations \eqref{411} to \eqref{413}
simplify significantly.  The result is
\begin{eqnarray}
    \det\left( D+\sumzero \rho^k N_k + \lambda \Omega - \lambda^2 I \right)
    & = &
    \lambda^4 + (\omega^2 - \alpha_{n/2 \pm 1}^2 + 2 \beta^2) \lambda^2
    \nonumber \\
    && \quad + 3 \omega^2 \left( \frac{1}{2} \alpha_{n/2 \pm 1} + \frac{3}{2}
                \alpha_{n/2}^2 - \beta^2 \right)
    \nonumber \\
    && \quad
             + \left( \frac{1}{2} \alpha_{n/2 \pm 1}^2 - \beta^2 \right)^2
         - \frac{9}{4} \alpha_{n/2}^4 .
    \label{401}
\end{eqnarray}
For a moment, let us write this biquadratic polynomial (in $\lambda$)
in a simple generic form and equate it to zero
\[
    \lambda^4 + A \lambda^2 + C = 0.
\]
The quadratic formula then tells us that
\[
    \lambda^2 = \frac{-A \pm \sqrt{A^2 - 4C}}{2} .
\]
To get the eigenvalues, we need to take square roots one more time.  The only
way for the resulting eigenvalues not to have
positive real part is for $\lambda^2$ to
be real and nonpositive.  Necessary and sufficient conditions for this are
that
\begin{eqnarray}
    A & \ge & 0 \label{1111} \\
    C & \ge & 0 \label{1112} \\
    A^2 - 4C & \ge & 0. \label{1113}
\end{eqnarray}
It turns out that the third condition implies the first two
(we leave verification of this fact to the reader).
In terms
of computable quantities, this third condition can be written,
after simplification, as
\[
    \omega^4
    + (-8 \alpha_{n/2 \pm 1}^2 - 18 \alpha_{n/2}^2 + 16 \beta^2) \omega^2
    + 9 \alpha_{n/2}^4 \ge 0 .
\]
Again we use the quadratic formula to find that
\[
    \omega^2 \ge
    4 \alpha_{n/2 \pm 1}^2 + 9 \alpha_{n/2}^2 - 8 \beta^2
    + \sqrt{
        (4 \alpha_{n/2 \pm 1}^2 + 9 \alpha_{n/2}^2 - 8 \beta^2)^2 - 9 \alpha_{n/2}^4
    }.
\]
or
\[
    \omega^2 \le
    4 \alpha_{n/2 \pm 1}^2 + 9 \alpha_{n/2}^2 - 8 \beta^2
    - \sqrt{
        (4 \alpha_{n/2 \pm 1}^2 + 9 \alpha_{n/2}^2 - 8 \beta^2)^2 - 9 \alpha_{n/2}^4
    }.
\]
It is the greater-than constraint that is relevant and so we take
the positive root. Finally, we recall that
\begin{eqnarray*}
    \omega^2 &=& \frac{GM}{r^3} + \frac{Gm}{r^3} \In \\
    \alpha_{n/2}^2 &=& \frac{Gm}{2r^3}(\Jn - \Jtnn) \\
    \beta^2 &=& \frac{Gm}{2r^3} \In
\end{eqnarray*}
and so the inequality on $\omega^2$ reduces to
\begin{eqnarray}
    \frac{M}{m}
    &\ge&
    2(\Jn-\Jtnnpm) + \frac{9}{2}(\Jn - \Jtnn) - 5 \In
    \nonumber \\
    && \quad + \sqrt{
        \left(2(\Jn-\Jtnnpm) + \frac{9}{2}(\Jn - \Jtnn) - 4 \In\right)^2
    - \frac{9}{4} \left( \Jn - \Jtnn \right)^2
    } .
    \label{402}
\end{eqnarray}

The second column in Table \ref{tab1} shows thresholds computed
using this inequality.  It is clear that for even values of $n$ greater than
$7$, this threshold matches the numerically derived threshold shown in the
first column in the table. This suggests that inequalities
analogous to \eqref{402} derived for $j \ne n/2$ are less
restrictive than \eqref{402}.
The proof of this statement is obviously more complicated than the $j = n/2$
case because the general case includes a linear term
($B_j \ne 0$) which vanishes in the $j = n/2$ case.  The linear
term makes it impossible simply to use the quadratic formula and
therefore any analysis involves a more general analysis of a
quartic equation.
\citet{1991CeMDA..51...83S} analyzed the general case.
Although their notations are different, the fundamental
quantities are the same and so their analysis is valid here as well.
Rather than repeating their complete analysis, we simply outline the basic
steps in the next subsection.

\subsubsection{The Subcase Where $j \ne n/2$.}

Let $\alpha_{j \pm 1}$ denote the average of $\alpha_{j+1}$ and
$\alpha_{j-1}$:
\[
    \alpha_{j \pm 1} = (\alpha_{j+1} + \alpha_{j-1})/2 .
\]
If we were to assume, incorrectly, for the moment that terms involving the
difference $\alpha_{j+1} - \alpha_{j-1}$ were not present in
\eqref{411}--\eqref{413}, then an analysis analogous to that given in the
previous subsection would give us the following inequality:
\[
    \omega^2 \ge
    4 \alpha_{j \pm 1}^2 + 9 \alpha_{j}^2 - 8 \beta^2
    + \sqrt{
        (4 \alpha_{j \pm 1}^2 + 9 \alpha_{j}^2 - 8 \beta^2)^2 - 9 \alpha_{j}^4
    }.
\]
Next, one uses the fact that $\alpha_{j \pm 1}^2$ is unimodal as a function of
$j$ taking its maximum value at $j=n/2$.  Hence, the inequality associated
with $j = n/2$ is the strictest of these inequalities.
Finally, the difference terms are treated as small perturbations to this
simple case and a homotopy analysis shows that the $j=n/2$ case remains the
strictest case even as the difference terms are fed in.

\subsubsection{Large $n$}

When $n$ is large, $\Jtnnpm \approx \Jtnn$ and $\Jn \gg \In$.
Furthermore,
\begin{eqnarray*}
    \Jtnn
    \approx
    \frac{1}{2} \sum_{k=1}^{n/2} \frac{(-1)^k}{\sin^3(k\pi/n)}
    & \approx &
    \frac{n^3}{2\pi^3} \sum_{k=1}^{\infty} \frac{(-1)^k}{k^3} \\
    & = &
    \frac{3}{4} \frac{n^3}{2\pi^3} \sum_{k=1}^{\infty} \frac{1}{k^3}
    \approx
    \frac{3}{4} \; \frac{1}{2} \sum_{k=1}^{n/2} \frac{1}{\sin^3(k\pi/n)}
    \approx
    \frac{3}{4} \Jn .
\end{eqnarray*}
Hence, \eqref{402} reduces to
%
%
\[
    \frac{M}{m} \ge \frac{7}{8} (13+2\sqrt{10}) \Jn,
\]
or, equivalently,
\begin{equation} \label{500}
    m \le \frac{M}{ \frac{7}{8} (13+2\sqrt{10}) J_n} \approx 2.299 M/n^3 ,
\end{equation}
%
which is precisely the answer Maxwell obtained 150
years ago. Of course, we have assumed here that $n$ is even. 
For the odd case, as
$n\rightarrow \infty$, $|\alpha_{j-1}-\alpha_{j+1}|\rightarrow 0$
so that the odd quartic equation for $j=(n-1)/2$  reduces to the
the even equation for $j=n/2$ giving the same stability criteria
as the even particle case. This can be seen in our simultions as
well. The case $n=100$ and $n=101$ give the same threshold to
several significant figures.
%
%
%
%

\subsection{The Case Where $j = 1$ Or $j = n-1$.} \label{subsec1}

By symmetry, these two cases are the same.  Hence, we consider only $j=1$.
Again after some manipulation in which we exploit the fact that
$\alpha_0^2 = 0$, we arrive at
\begin{equation}
    \det\left( D+\sumzero \rho^k N_k + \lambda \Omega - \lambda^2 I \right)
    =
    \lambda^4 + A_j \lambda^2 + iB_j \lambda + C_j = 0,
    \label{414}
\end{equation}
where
\begin{eqnarray}
    A_1 & = & \omega^2
             - \frac{1}{2} \left( n\gamma + \alpha_{2}^2 \right)
         + 2 \beta^2 \label{415} \\
    B_1 & = & - \omega \left( n\gamma - \alpha_{2}^2 \right)
            \label{416} \\
    C_1 & = & 3 \omega^2 \left( \frac{1}{4} \alpha_2^2
                  + \frac{3}{2} \alpha_1^2 - \beta^2
                  - \frac{1}{2} n\gamma
                         \right)
           + \left( \frac{1}{4} ( n\gamma + \alpha_{2}^2 )
                - \beta^2 \right)^2
    \nonumber \\ &   & \quad
            - \frac{1}{16} (n\gamma-\alpha_{2}^2)^2
                    - \frac{9}{4} \alpha_1^2 (\alpha_1^2-n\gamma)
           , \label{417}
\end{eqnarray}
and
\begin{equation} \label{418}
    \gamma = \frac{Gm}{r^3} .
\end{equation}

Note that the coefficient $B_1$ is imaginary whereas the other three
coefficients are real.  This suggests making the substitution $\mu = i
\lambda$.  In terms of $\mu$, \eqref{414} becomes a quartic equation with all
real coefficients:
\begin{equation} \label{602}
    \mu^4 - A_1 \mu^2 + B_1 \mu + C_1 = 0.
\end{equation}
This equation either has four real roots or not.  If it does, then the
corresponding values for $\lambda$ are purely imaginary and the system could
be stable.  If, on the other hand, there are two or fewer real roots,
then at least one pair of roots to \eqref{602}
form a conjugate pair and therefore
the corresponding pair of values for $\lambda$ will
be such that one has positive real part and the other negative real part.
Hence, in that case, the system is demonstrably unstable.
Simple numerical investigation reveals that this is precisely what happens
when $2 \le n \le 6$ regardless of the mass ratio $M/m$.
\\%
To see why, let us consider just the case when $M/m$ is very large and hence
the ratio
\[
    r = m/M
\]
is very close to zero.  In this asymptotic regime,
\begin{eqnarray*}
    A_1 & = & a M \\
    B_1 & = & b \sqrt{M} m \\
    C_1 & = & c M m ,
\end{eqnarray*}
where $a > 0$ and the sign of $c$ is the same as the sign of
$ \frac{1}{4} \alpha_2^2 + \frac{3}{2} \alpha_1^2 - \beta^2 - \frac{1}{2} n\gamma $.
Substituting $rM$ for $m$ and making the change of variables defined by
$\nu = \mu/\sqrt{M}$, we get
\[
    \nu^4 - a\nu^2 + br \nu + cr = 0 .
\]
For $r=0$, this equation reduces to $\nu^4 - a \nu^2 = 0$ which has three real
roots, a positive one, a negative one, and a root of multiplicity two at
$\nu = 0$.  By continuity, for $r$ small but nonzero, the quartic still has
a positive root and a negative root but the double root at $\nu = 0$ can
either disappear or split into a pair of real roots.  Since this bifurcation
takes place in a neighborhood of the origin, the quartic term can be ignored
and the equation in a tiny neighborhood of zero reduces to a quadratic
equation:
\[
    - a\nu^2 + br \nu + cr = 0 .
\]
This equation has two real roots if and only if its discriminant is
nonnegative:
\[
    r (r b^2 + 4  a c) \ge 0 .
\]
Hence, if $c$ is negative there will not be a full set of real roots for $r$
very small and hence the ring system will be unstable in that case.  In other
words, the system will be unstable if
\[
    \frac{1}{4} \alpha_2^2 + \frac{3}{2} \alpha_1^2 - \beta^2 - \frac{1}{2} n\gamma < 0.
\]
This equation reduces to
\begin{equation} \label{1504}
     \sum_{k=1}^{n-1}\frac{1}{\sin\left(\frac{\pi k}{n}\right)}
	    -n-\frac{1}{2}\cot\left(\frac{\pi}{2 n}\right)
                         < 0 .
\end{equation}
It is easy to check that the expression is negative for $n=2,\ldots,6$ and
positive for $n \ge 7$.  Therefore, we have proved that ring systems are
unstable for $n=2,\ldots,6$ at least when $m$ is very small relative to $M$.
We have not proved the result for larger values of $m$ but it seems that such
a case should be even more unstable, which is certainly verified by our
simulator.

\section{Ring Density}
Suppose that the linear density of the boulders is $\lambda$.  That is,
$\lambda$ is the ratio of the diameter of one boulder to the separation
between the centers of two adjacent boulders.  Then the diameter of a single
boulder is $\lambda (2 \pi r/n)$.  Hence, the volume of a single boulder is
$(4 \pi/3)(\lambda \pi r / n)^3$.  Let $\delta$ denote the density of a
boulder.  Then the mass of a single boulder is
$(4 \pi/3)(\lambda \pi r / n)^3 \delta$.
%
If we assume that the density of Earth is about $8$
times that of a boulder (Earth's density is $5.5$ and Saturn's
moons have a density of about $0.7$ being composed of porous
water-ice), then we have
\[
    \delta = \frac{1}{8} \frac{m_E}{(4 \pi/3) r_E^3} ,
\]
where $m_E$ denotes the mass of Earth and $r_E$ denotes its radius.
Combining all of these factors and assuming the central mass is equal to
Saturn's mass and the ring's radius is about the radius of the Cassini
division ($120,000$km), we see that the upper bound on the linear
density of boulders is
\[
    \lambda \le \left( 8 \frac{M/m_E}{(25.65)(0.01938)} \right)^{1/3}
            \frac{r_E}{r} = 0.219.
\]
In other words, the linear density cannot exceed $22\%$ otherwise the ring
will be unstable.  Of course, this is for a one dimensional circular ring of
ice boulders.
Analysis of a two dimensional annulus or the full three dimensional
case is naturally more complicated.  Nonetheless the $22\%$ linear density
figure matches surprisingly well with the measured optical density which
hovers around $0.05$ to $2.5$.

\section{Numerical Results}

We have computed stability thresholds three different ways for various
finite $n$.

First, we numerically solved for all eigenvalues of the $4x \times 4n$ matrix
in \eqref{101} and did a binary search to locate the smallest mass ratio $M/m$
for which no eigenvalue has positive real part.  We then tranlated this
threshold into a value of $\gamma$ for the threshold expressed as
\[
    m \le \gamma M/n^3
\]
and tabulated those results in the column labeled {\em Numerical} in Table
\ref{tab1}.

Secondly, for even values of $n$ we used equation \eqref{402} to
derive $\gamma$ threshold values.  These values are reported in the second
column of thresholds in Table \ref{tab1}.  Note that for even values of $n$
larger than $7$, these
results agree with those obtained numerically.

Lastly, and perhaps most interestingly, the third column of results in
the table are stability thresholds that were estimated using a
simulator (\cite{VanRingsApplet})
based on a leap-frog integrator (\cite{ST94,HMM94}).
In this column, two values are given.
For the larger value instability has been decisively observed.
However, verifying stability is more challenging since one should in principle
run the simulator forever.  Rather than waiting that long, we use the rule of
thumb that if the system appears intact for a period of time ten times greater
than the time it took to demonstrate instability, then we deem the system
stable at that mass.  This is how the lower bounds in the table were obtained.
It is our belief that a good (simplectic) simulator provides the most
convincing method to discriminate between stable and unstable orbits at least
when the number of bodies remains relatively small, say up to a few hundred.
Unstable orbits reveal themselves quickly as the initial inaccuracy of double
precision arithmetic quickly cascades into something dramatic if the system is
unstable.  If, on the other hand, the system is stable then the initial
imprecisions simply result in an orbit that it close to but not identical to
the intended orbit.  The situation does not decay.  Any reader who has never
experimented with a good simplectic integrator is strongly encouraged to
experiment with the Java applet posted at
\begin{quote}
\url{http://www.princeton.edu/~rvdb/JAVA/astro/galaxy/StableRings.html}
\end{quote}
as hands on experience can be very convincing.

Of course, the amazing thing about the simulator results is that they match
the numerical results in the first column.  The thresholds determined by
linear stability analysis only tell us definitively that for $m$ larger than
the threshold, the system is necessarily unstable.  But, for $m$ smaller than
the threshold, the mathematical/numerical analysis says nothing since in
those cases, the eigenvalues are all purely imaginary.  Yet, simulation
confirms that the thresholds we have derived are truely necessary and
sufficient conditions for stability.

As shown in Section \ref{subsec1} for $2 \le n \le 6$, the system is unstable.
The simulator verifies this.
In these cases, there is lots of room to roam before one body
catches up to another.  Even the tiniest masses are unstable.
The case $n=2$ is especially interesting.  For this
case, the two small bodies are at opposite sides of a common orbit.
Essentially they are in L3 position with respect to each other.  For the
restricted 3-body problem (where one body has mass zero) it is well-known that
L3 is unstable no matter what the mass ratio.  Our results bear
this out.

\begin{table}
\begin{center}
\begin{tabular}{r|lll}
      & \multicolumn{3}{|c}{Threshold Value for $\gamma$} \\[0.1in]
  $n$ & Numerical & Eq. \eqref{402} & Simulator \\
      &           & (even $n$ only)      &           \\\hline
   2 & 0.000 & 4     & [0.0, 0.007] \\
   6 & 0.000 & 2.487 & [0.0, 0.025] \\
   7 & 2.452 & -     & [2.45, 2.46] \\
   8 & 2.412 & 2.4121 & [2.41, 2.42] \\
  10 & 2.375 & 2.3753 & [2.37, 2.38] \\
  36 & 2.306 & 2.3066 & [2.30, 2.35] \\
 100 & 2.300 & 2.2999 & [2.30, 2.31] \\
 101 & 2.300 & -      & [2.30, 2.31]
\end{tabular}
\caption{Estimates of the stability threshold (i.e., $\gamma$ in an inequality
of the type $m \le \gamma M/n^3$).  The first column contains numerically
derived obtained by a brute-force
computation of the eigenvalues together with a
simple binary search to find the first point is at which an
eigenvalue takes on a positive real part.
The second column gives thresholds computed using \eqref{402}.
The column of {\em simulator}
values corresponds to results from running a leap-frog integrator and noting
the smallest value of $\gamma$ for which instability is clearly demonstrated.
This is the larger of the pair of values shown.  The smaller value is a nearby
value for which the simulator was run ten times longer with no overt
indication of instability.}
\label{tab1}
\end{center}
\end{table}

\bibliography{../lib/refs}   
\bibliographystyle{plainnat}    

\end{document}